\documentstyle[12pt,epsf]{article}

\begin{document}
\setlength{\unitlength}{1mm}
\textwidth 15.0 true cm 

\headheight 0 cm
\headsep 0 cm
\topmargin 0.4 true in
\oddsidemargin 0.25 true in
\input epsf

\newcommand{\beq}{\begin{equation}}
\newcommand{\eeq}{\end{equation}}
\newcommand{\be}{\begin{eqnarray}}
\newcommand{\ee}{\end{eqnarray}}
\renewcommand{\vec}[1]{{\bf #1}}

\newcommand{\grpicture}[1]
{
    \begin{center}
        \epsfxsize=200pt
        \epsfysize=0pt
        \vspace{-5mm}
        \parbox{\epsfxsize}{\epsffile{#1.eps}}
        \vspace{5mm}
    \end{center}
}

\begin{flushright}

SUBATECH--00--20\\
ITEP/TH-54/00\\

\end{flushright}

\vspace{0.5cm}

\begin{center}

{\Large\bf  Classical Yang-Mills Vacua on $T^{3}$ : Explicit Constructions.}

\bigskip

{\bf  K.G. Selivanov}\protect\(^{1}\protect \) and   {\bf A.V. Smilga}\protect\( ^{2,1}\protect \)
\\

\bigskip

 \protect\( ^{1}\protect \) {\it ITEP, B. Cheremushkinskaya 25, 
117218,  Moscow, Russia.}\\
 \protect\( ^{2}\protect \){\it SUBATECH, Universit\'e de
Nantes,  4 rue Alfred Kastler, BP 20722, Nantes  44307, France. }\\

\end{center}

\bigskip

\begin{abstract}

Flat connections for unitary gauge  groups on a 3--torus with twisted 
boundary conditions as well as recently discovered periodic 
nontrivial flat 
connections  with  ``nondiagonalizable'' triples  of holonomies for higher 
orthogonal and exceptional 
groups are constructed explicitly in terms of 
Jacobi theta functions with rational characteristics. The (fractional)
Chern-Simons numbers of these vacuum gauge field configurations are
verified by direct computation. 

\end{abstract}

\section{Introduction}

Gauge theories on a torus have been studied for  a long 
time \cite{thooft}.
The main reason why this subject is interesting   is that the torus provides a natural infrared cutoff  which does not break translational invariance and supersymmetry. The Euclidean 4--dimensional torus is used in lattice calculations. If we stay in the Hamiltonian framework,  only space is compactified and the theory is defined on $T^3 \times R$. 
For small spatial tori the effective coupling constant is also small
and  the vacuum structure of the theory can be studied 
in the Born--Oppenheimer framework. This approach is especially useful in the supersymmetric case, where the number of exact vacuum states does not depend on the size of the torus \cite{witten}.
  It was also suggested that the modes which are most relevant in the strong coupling regime are topologically distinguished on $T^{3}$
 \cite{thooft} .

One of the important technical aspects is that  gauge theories on a torus admit 
nontrivial {\it twisted} boundary conditions \cite{thooft}. 
\footnote{This is so if the theory does not involve the fields in the fundamental or other
representation for which the group of the center acts faithfully.  For example, the 
twisted boundary conditions are not admissible for standard QCD involving quarks.} 
They have the form 
\begin{eqnarray}
\label{twist}
A_i(x+1,y,z) \ &=&\ PA_i(x,y,z)P^{-1}\ ,\nonumber\\
A_i(x,y+1,z)\ &=&\ QA_i(x,y,z)Q^{-1}\ ,\\
A_i(x,y,z+1)\ &=&\ SA_i(x,y,z)S^{-1}\ ,\nonumber
\end{eqnarray}
 where $i = 1,2,3$ and the periods of the torus are normalized to 1. Now, $P$ , $Q$ , 
and $S$ are constant elements of the
gauge group forming  so called Heisenberg pairs:
\beq
\label{pair}
QP=\omega_1 PQ\ ,\ \  QS=\omega_2 SQ\ ,\ \ SP=\omega_3 PS\ ,
\eeq
with $\omega_i$ belonging to the center of the  group. If at least one of $\omega_i$ 
is a nontrivial element of the center, 
 the conditions (\ref{twist},\ \ref{pair}) mean from the mathematical standpoint that 
our fiber bundle is topologically nontrivial (not reduced to the direct product 
$G \times T^3$). 

To study the vacuum structure of quantum Yang--Mills theory, one has first to understand 
the structure of classical vacua. The latter are given by  gauge field configurations with 
zero field strength,  flat connections in mathematical language. Classification of all flat 
connections on a 3--torus  is an  interesting and nontrivial mathematical problem.  For 
unitary groups it was largely  solved in Refs.[1,2].  It turns out that any topologically 
trivial flat connection $A_i(x,y,z)$ can be gauge--transformed to constant commuting $A_i$. 
 A distinct vacuum is characterized by a set of holonomies $\Omega_i  = \exp\{iA_i\}$, 
with each holonomy lying on the maximal torus of the group.

When twist is allowed, the moduli space of classical vacua contains generically several 
disconnected components. Consider the case $S=1$, which implies $\omega_2 = \omega_3 = 1$, 
and assume that $\omega_1$ is the primitive N-th root of unity  $\epsilon = \exp\{2\pi i/N\}$  
(or some other element generating the whole center subgroup $Z_N$). In this case,
the moduli space of vacua factorized over all gauge transformations, including the gauge 
transformations of  ``instanton nature"  which change the Chern--Simons (CS) number of a 
gauge field configuration by an integer, contains just $N$ isolated points. 
 The  CS  numbers of these isolated vacua are fractional 
$ N_{CS} = {p}/{N}$, $p = 0,\ldots, N-1$. We will call the connections with fractional
CS number {\it interesting}. 
\footnote{The word ``nontrivial'' is too broad and vague while 
``topologically nontrivial'' would not be  correct since 
the topologically nontrivial connections
of instanton nature are not interesting in the present context,
while certain topologically trivial (periodic) connections are. }  

If  $\omega_1 = \epsilon^k$, where $k$ is a divisor of $N$, the moduli space contains
$N/k$ disconnected components. In each such component the holonomies $\Omega_i$ lie on a subtorus 
of dimension $k-1$ of the maximal  torus .  The  corresponding CS  numbers   are multiple integers 
of $ {k}/{N}$. 

It can be shown \cite{Baal1} that the boundary conditions (\ref{twist})  with nontrivial
 $S$ bring about  nothing new and the 
problem is always reduced to one of the cases described above.

For other groups the situation is more complicated, and the problem was solved only recently. 
It turned out that, for higher orthogonal and exceptional groups,  the moduli space of 
classical
 vacua involves disconnected components even in the case of trivial twists \cite{witten2,krs}.
(For symplectic groups with trivial twist, there is only one component. This case was
analyzed back in Ref.\cite{witten}.) 
 The complete classification of periodic flat connections for an arbitrary gauge group was 
constructed in \cite{ks,k}.  With the classical vacuum moduli space in hand,
also quantum problem can be solved. In all cases, the number of
quantum vacuum state in  
pure ${\cal N} = 1$ supersymmetric Yang--Mills theory coincides with the so called dual 
Coxeter number $h^\lor$ (or just the adjoint Casimir eigenvalue $c_V$) of the group.
\footnote{For theories involving extra matter supermultiplets it may be completely different \cite{extravac}.}
 The 
classification of flat connections with nontrivial twist for nonunitary groups was 
constructed in \cite{bfm} (for symplectic and orthogonal groups it was pedagogically 
explained in the last section of recent \cite{witten3}). Again, the number of quantum vacuum 
states always coincides with the dual Coxeter number {\it independently} of the boundary 
conditions chosen. 

In all these more complicated cases,  the basic building block used to construct   nontrivial
 flat connections are flat connections for unitary groups with twisted boundary conditions. 
Let us recall how it is done for periodic connections. Any flat connection is characterized 
by a triple of commuting holonomies. In contrast to the simple case of unitary groups, such a 
commuting triple is not always ``diagonalizable'', i.e. it 
cannot be always conjugated (gauge transformed)  to the maximal torus (to be precise: 
conjugated to a triple belonging to the maximal torus ). A 
generic such nondiagonalizable commuting triple 
(an {\it exceptional } triple as defined
in Ref. \cite{ks}) is constructed as follows. One chooses the first holonomy 
of the triple $\Omega_1$ in such a 
way that its centralizer (a subgroup of the large group $G$ 
containing all elements of $G$ commuting with $\Omega_1$) 
is a group whose semi-simple component is a direct product of several $SU(N_{i})$  groups 
factorized 
over a subgroup of its center $\prod_i Z_{N_i}$. This factorization results in that the 
fundamental group of the centralizer 
involves a finite  subgroup as a factor (the elements whose centralizer enjoys this property
are called  {\it exceptional}) and this allows one to pick up two commuting elements 
$\Omega_{2}, \Omega_{3}$ in the centralizer  which cannot be conjugated to its maximal torus. 
Then the triple $\Omega_{1}, \Omega_{2}, \Omega_{3}$ 
cannot be conjugated to the maximal torus in $G$. 

Consider as a simplest  example the  group $G_2$. It involves a unique (up to conjugation) 
{ exceptional}  element whose centralizer  is $[SU(2) \times SU(2)]/Z_2$, where the 
factorization 
is done over the diagonal subgroup of the center $Z_2 \times Z_2$
of $SU(2) \times SU(2)$ ( i.e. the element $(-1, -1)$ of  $SU(2) \times SU(2)$
is identified with 1 ).  Then one can choose 
the elements $\Omega_{2,3}$ in the centralizer in such a way that their liftings 
in $SU(2) \times
SU(2)$ are 
$\tilde \Omega_2 =\ (P, P),\  \tilde\Omega_3 =\ (Q, Q)$ 
(in obvious  notations corresponding to the direct product structure) such that $P,Q$ form a Heisenberg pair, 
$PQ = -QP$, in each $SU(2)$ component. Factorization over the diagonal $Z_2$ makes the pair 
$\Omega_{2}, \Omega_{3}$ commuting. As a noncommuting Heisenberg pair cannot obviously be 
conjugated to a maximal torus in $SU(2)$, the pair $\Omega_{2}, \Omega_{3}$
cannot be conjugated to the maximal torus in $[SU(2) \times SU(2)]/Z_2$ and the whole triple 
cannot be conjugated to the maximal torus
in $G_2$.  Note that any two
of the holonomies can be conjugated to a maximal torus in $G_2$. But the corresponding tori for
the subsets $\Omega_{1,2},\ \Omega_{1,3}$, and $\Omega_{2,3}$ are different.

A similar construction works also in all other cases. It is more or less
clear that a periodic flat connection based on an exceptional commuting
triple of holonomies can be constructed if a flat connection in $SU(N)$,
with holonomies  $\Omega_{2,3} \ =\  P, Q$ such that $P,Q$ form a Heisenberg pair  and $\Omega_1$ belonging to the centralizer of $P,Q$ in $SU(N)$, is known. In 
particular, the CS number of the former coincides with that of the latter and is, generically, 
fractional. A fractional CS number of new nontrivial vacua is
their important property, which allows one to ascribe to the corresponding quantum vacua correct 
fermionic charges and make contact with 
what is known about the vacuum structure of supersymmetric Yang--Mills theory in  large volume 
\cite{witten3}  
 
The existence of the twisted gauge field configurations with fractional CS number was known before. 
Their numerical study was
performed in \cite{simulation}. The main goal of our paper is to present simple explicit analytic 
formulae for these configurations. 

In sect. 2 we construct a flat connection corresponding to holonomies
forming the Heisenberg
pair in $SU(N)$. The result is expressed 
in terms of Jacobi $\Theta$ functions with rational characteristics.  
 \footnote{The analogies with the problem of motion 
of a charged particle in a constant homogeneous magnetic field on a
2D torus are very instructive here.}
We perform a direct computation of  the 
CS number of the gauge field configuration thus obtained. 
 We confirm that $N_{CS}$ is a multiple integer of $k/N$ for 
$\omega_1 = \epsilon^k$ with integer $N/k$ .

In sect. 3 we lift the gauge fields associated with the  $SU(N)$
Heisenberg pairs to  gauge fields with exceptional triples of commuting 
holonomies
in the orthogonal and exceptional groups (a different construction for the 
flat periodic 
connections in $Spin(7)$ has been done in \cite{s} ). The computation of 
the CS number is  
reduced to the previous case. 
  The same can be  done for twisted connections in nonunitary groups. 
We discuss the simplest such case, which is $Sp(4)$.  

In the last section we present our conclusions and 
prospects for future research.

\section{Twisted  flat connections in $SU(N)$.}

Let us  first  assume that $\omega_1 = \epsilon$.
The zero curvature gauge fields can be represented in the form
\beq
\label{flat}
A_{i}=U^{-1} \partial_{i} U\ .
\eeq
Following \cite{thooft,witten}, we search for a  gauge group matrix $U(x,y,z)$ obeying the  
boundary conditions:
\begin{eqnarray}
   \label{twistu}
 U(x+1)=PU(x)P^{-1}\nonumber\\
 U(y+1)=QU(y)Q^{-1}\\
 U(z+1)=\epsilon U(z)\ ,\nonumber
  \end{eqnarray}
where the dependence on ``irrelevant" variables ($y,z$ in the first line, etc. ) 
is not displayed. Apparently, the conditions (\ref{twistu}) are compatible with the 
conditions for the gauge fields in Eq. (\ref{twist}). 

We start our construction of the $SU(N)$ gauge field obeying 
Eqs.(\ref{flat}), (\ref{twistu})  with the ansatz
\beq
\label{ansatz}
U\ =\ e^{{2\pi i z} T(x,y)},
\eeq
where $T(x,y)$ is a Hermitian  $su(N)$ matrix conjugated to the matrix
\footnote{In fact, it is one of the {\it fundamental coweights}.}
   \beq
 \label{t8}
 T_0 \ =\ \frac 1N {\rm diag} (1, \ldots, 1, 1-N)
   \eeq
Apparently, $U|_{z=0}=1$ and  $U|_{z=1}= \epsilon$, so the third 
condition of Eq. (\ref{twistu}) is satisfied. The other two conditions 
translate as the following conditions on  $T(x,y)$:
\begin{eqnarray}
\label{twistt}
T(x+1)=PT(x)P^{-1}\nonumber\\
T(y+1)=QT(y)Q^{-1}\ .
\end{eqnarray}
It is rather easy to satisfy these conditions for $SU(2)$. If 
$P = i\sigma_3$ and $Q = i\sigma_1$ 
(any Heisenberg pair in $SU(2)$ can be conjugated to this form) , 
one can choose
  \be
  \label{txy}
T(x,y) \ =\ \frac 12 \ \frac {\sigma_1 \cos (\pi x) +  \sigma_3 \cos (\pi y) +
\sigma_2 \cos [\pi (x+y)]}{\sqrt{\cos^2 (\pi x) + \cos^2 (\pi y) + 
\cos^2 [\pi (x+y)]}} \ ,
   \ee
 where the square root factor is inserted for  proper normalization.
  A similar in spirit formula was written in Ref.\cite{simulation}{\it a}  
for the case   $P = i\sigma_1$, 
$Q = i\sigma_2$, $S= i\sigma_3$. It is difficult, however, to generalize the solution (\ref{txy}) 
to the case of higher $N$.
 
To solve (\ref{twistt}) for arbitrary $N$, we first notice that the matrices conjugated to $T_0$ form 
$CP^{N-1}=\frac{SU(N)}{SU(N-1) \times U(1)}$ orbit of $SU(N)$. They are 
conveniently parameterized as follows
\beq
\label{fund}
T_{ij}(x,y)\ =\ \frac 1N \delta_{ij}- \psi_{i}(x,y) {\psi}^\dagger_{j}(x,y),
\eeq
where $\psi_{i}$ is a $N$-component complex column normalized to unity: 
\beq
\label{norm}
{\psi}^\dagger \psi=1.
\eeq
Now, $ \psi$ is an element 
of the fundamental representation of $SU(N)$, and the parameterization
(\ref{fund}) of the orbit $\frac{SU(N)}{SU(N-1) \times U(1)}$  may be called 
{\it fundamentalization}. A traceless  Hermitian matrix $T(x,y)$ from 
Eq. (\ref{fund}) has $2N-2$ 
real parameters 
[$N$ complex parameters in the column $\psi_{i}$ minus 1 real parameter for 
the normalization 
Eq. (\ref{norm}) and minus 1 real parameter for the  irrelevant common phase of
$\psi_{i}$ in Eq. (\ref{fund})], which is equal to  the dimension of
  $\frac{SU(N)}{SU(N-1) \times U(1)}$ space. 

The boundary condition  (\ref{twistt}) is reduced to
\begin{eqnarray}
\label{twistp}
\psi(x+1)\ &=&\ e^{i\alpha(x,y)} P \psi(x)\nonumber\\
\psi(y+1)\ &=&\ e^{i\beta(x,y)} Q \psi(y)\ ,
\end{eqnarray}
where  real functions $\alpha(x,y)$ and $\beta(x,y)$ should be chosen 
 to compensate the nontrivial commutant (\ref{pair}) of $P$ and $Q$ and to  make $\psi(x+1,y+1)$ uniquely defined. The latter self--consistency condition  implies
\beq
\label{coc}
e^{-i\alpha(x,y)}e^{-i\beta(x+1,y)}e^{i\alpha(x,y+1)}e^{i\beta(x,y)}\ =\ \omega_1 \ =\ 
\epsilon\ ,
\eeq
and we make a choice 
\beq
\label{gauge}
\alpha(x,y)\ =\ \frac{2\pi   y}{N},\; \ \ \beta(x,y)\ =\ 0 \ .
\eeq
 The phases $\alpha(x,y), \beta(x,y)$ can be interpreted as vector potentials $A_{x,y}$ of  
an auxiliary constant Abelian magnetic field with flux $\Phi = 1/N$ on the 2--torus
(and their exponentials are the
corresponding abelian holonomies).

Being expressed in words, Eq. (\ref{twistp})  means that we need to construct a global section 
of a  $\frac{SU(N) \times U(1)}{Z_{N}}=U(N) $ bundle over $T^{2}$ with $C^{N}$
as a typical fiber. $e^{i\alpha(x,y)} P$ and $e^{i\beta(x,y)} Q$
are the transition matrices. 
The first Chern class of the bundle is 
  \be
  \label{Chern1}
  c_1 \ =\ \frac 1{2\pi} \int \ {\rm Tr} \{F \} \ =\ N\Phi \ =\ 1 \ .
 \ee
 This is a problem which the Jacobi
$\Theta$ functions with rational characteristics (see Ref. \cite{mum} for definitions and 
notations) are tailor--made for.

It was shown in Ref. \cite{thooft} that any Heisenberg pair in $U(N)$ 
satisfying 
$QP = \epsilon PQ$ can be conjugated to 
  \be
  \label{PQ}
 P= \ e^{i\delta_P}\left( \begin{array}{lllcl}
1 & 0 &   0 & 0 & \ldots  \\
0 & \epsilon &  0 & 0 &  \ldots  \\
0 & 0 & \epsilon^{2} & 0 &   \ldots  \\
{\vdots} & {\vdots} & {\vdots} & {\vdots} & {\vdots}  \end{array} \right) \ , 
 \nonumber \\
Q= \ e^{i\delta_Q}\left( \begin{array}{lllcl}
0 & 1 &   0 & 0 & \ldots  \\
0 &  0 &  1 & 0 &  \ldots  \\
0 & 0 & 0 & 1 &   \ldots  \\
{\vdots} & {\vdots} & {\vdots} & {\vdots} & {\vdots}\\
1 & 0 & 0 & 0 & {\ldots}  \end{array} \right)  \ .
  \ee
  By choosing some special $\delta_{P,Q}$, one can also 
write a 
``canonical" Heisenberg pair for $SU(N)$, but we do not need this because  an overall phase 
of $P,Q$ is not relevant in Eq. (\ref{twistt}). 

Notice that $Q$ acts on the column $\psi$ by
cyclically shifting its elements one step up so that the second 
condition in Eq. (\ref{twistp}) simply fixes all the components
  $\psi_{j}$ in terms of  $\psi_{1}$,
\beq
\label{pj}
\psi_{1+j}(x,y)=\psi_{1}(x,y+j)
\eeq
and  requires thereby periodicity of  $\psi_{1}$ when $y$ is shifted by $N$,
\beq
\label{p1}
\psi_{1}(x,y+N)=\psi_{1}(x,y)\ .
\eeq
All other components $\psi_j$ also enjoy this property.
In view of Eqs.(\ref{pj} , \ref{PQ}, \ref{gauge}),  
 the first  condition in Eq. (\ref{twistp}) is reduced to
\beq
\label{p1x}
\psi_{1}(x+1,y)=e^{\frac{2\pi i y}{N}}\psi_{1}(x,y)\ .
\eeq

The conditions (\ref{p1}, \ref{p1x})  are obviously satisfied upon the choice
\beq
\label{p1t}
\psi_{1}(x,y)={\cal N}(x,y)\sum_{n \in Z} e^{-\pi (n+\frac{y}{N})^{2} + 2\pi i x(n+\frac{y}{N})},
\eeq
where ${\cal N}(x,y)$ is a periodical function of $x$ and $y$ with period 1. 
Other $\psi_{j}$ are defined via Eq. (\ref{pj}). The factor ${\cal N}(x,y)$ should be chosen such 
that the normalization condition (\ref{norm}) is satisfied .  
For  ${\cal N}$
to be well defined we need to check that $\psi_{j}$ do not have a common 
zero. To this end it is convenient to express $\psi_{j}$ in terms of
Jacobi $\Theta$ functions.
Using the definition of the theta functions 
$\Theta_{l/N, m/N}(z, \tau)$ with rational 
characteristics $l/N, \; m/N$ (see \cite{mum}),
\beq
\label{Theta}
\Theta_{l/N, m/N}(z, \tau)=\sum_{n \in Z} e^{ i\pi \tau (n+\frac{l}{N})^{2} +
2\pi i (n+\frac{l}{N})(z+\frac{m}{N})},
\eeq
one straightforwardly verifies that
\beq
\label{psi}
\psi_{j}(x,y)={\cal N}(x,y) e^{-\pi (\frac{y}{N})^{2}+2\pi i x\frac{y}{N}}
\Theta_{(j-1)/N, 0}\left(x+ i \frac{y}{N} , i\right).
\eeq
Now, $\Theta_{l/N, m/N}(z, \tau)$ have zeros at 
$z=(l/N+p+1/2) \tau + (m/N+q+1/2), \; p,q \in Z$, so $\psi_{j}(x,y)$
have no common zero. The factor ${\cal N}$ can thus be determined as 
\beq
\label{nice}
{\cal N}(x,y) = 
\frac{e^{\pi (\frac{y}{N})^{2}}}
{\sqrt{\sum_{l=0}^{N-1} |\Theta_{l/N, 0}(x+ i \frac{y}{N} , i)|^{2}}  } .          
\eeq

A digression is in order here.
The conditions (\ref{p1}, \ref{p1x}) are the same as the conditions imposed on a charged
 particle moving on the ``large"  2--torus (with 
$0 \leq x \leq 1,\ 0 \leq y \leq N$ ) in an external homogeneous magnetic 
field $B = 2\pi/N$ .  
It is known that such a problem is self--consistent only if
the total flux $\Phi = B{\cal A}/(2\pi)$ is integer. In our case, 
$\Phi_{\rm large \ torus} \ =\ 1$. Charged particle on a torus
in an arbitrary magnetic field with a given flux was considered
in \cite{DN}. The solution of the problem for homogeneous field allows one to calculate the functional integral in Schwinger model in topologically nontrivial sectors \cite{Wipf}.

If going back to the original ``small" torus $0 \leq x,y \leq 1$, we see a 
system of $N$ 
charged particles moving in a magnetic field of flux
$\Phi_{\rm small \ torus} \ =\ 1/N$. Fractional fluxes are now admissible 
because the corresponding 
wave functions 
satisfy the {\it flavor--twisted} boundary conditions (\ref{pj}). Such 
conditions were studied 
before in Ref. \cite{frac}. In multiflavor Schwinger model they allow 
for the presence of 
Euclidean gauge field configurations with fractional instanton number.

Note that, though the boundary conditions in our problem 
and in the problem of motion in an external magnetic field are
identical,  the functions (\ref{p1t}, \ref{psi})
{\it do} not solve the corresponding Schr\"odinger equation. The normalization condition (\ref{norm}) which we have to satisfy at  
{\it  every} point on the torus is alien to the Sturm--Liouville settings.

Substituting Eq. (\ref{psi}) into Eqs.(\ref{fund}), (\ref{ansatz}), 
and (\ref{flat}), we obtain the twisted connection we were seeking for.
All other solutions to the boundary conditions (\ref{twistu}) are related 
to this particular solution  under gauge transformations, including 
transformations of the instanton type.

Let us now compute the CS number of this field. We have
\beq
\label{cs}
N_{CS}\ =\ \frac{1}{8 \pi^{2}} \int_{T^{3}}  \ {\rm Tr}\left(AdA+
\frac{2}{3}A^{3}\right)
\eeq
which is normalized so that, on two flat gauge fields
related by an instanton, the CS number differs by 1.
Since the connection is flat, we actually
need to compute the integral

  \beq
 \label{cs1}
  N_{CS}\ =\ -\frac{1}{8 \pi^{2}}  \int d \vec{x}  
  \ {\rm Tr} \left\{ (\partial_{x} U^{-1}  \partial_{y}U 
  - \partial_{y} U^{-1}  \partial_{x}U ) 
  U^{-1} \partial_{z}U \right\}\ .
   \eeq
When the spatial manifold is $S^3$, Eq. (\ref{cs1}) defines 
$\pi_3[SU(N)]$  and is integer. The same holds for the normal untwisted torus. 
But in the twisted case the situation is different.

To find (\ref{cs1}), we substitute there $U(x,y,z)$ in the form   
(\ref{ansatz}),  (\ref{fund}).  
Then $U^{-1} \partial_{z}U\ =\ 2\pi i T(x,y)$. To 
find  the factors $\partial_{x,y} U^{-1}$,
$\partial_{x,y} U$ ,
it is convenient to represent  $U$ and  $U^{-1}$ as follows
\begin{eqnarray}
U=e^{\frac{2\pi i z}{N}}[1+(e^{-2\pi i z}-1) \Pi]
\nonumber\\
U^{-1}=e^{-\frac{2\pi i z}{N}}[1+(e^{2\pi i z}-1) \Pi]\ ,
\end{eqnarray}
with $\Pi_{ij} = \psi_i \psi^\dagger_j$, $\Pi^2 = \Pi$. 
Then Eq. (\ref{cs1}) is reduced to 
\begin{eqnarray}
\label{cs2}
N_{CS} \ =\ \frac{1}{\pi i}  \int dxdydz \sin^{2}(\pi z) 
{\rm Tr} \left\{  [(\partial_{x} \Pi )(\partial_{y} \Pi) 
-  (\partial_{y} \Pi )(\partial_{x} \Pi) ] \Pi \right\} \ =\ \nonumber \\
=\frac{1}{2 \pi i}  \int dxdy [\partial_{x}( \psi^\dagger \partial_{y}\psi )-
\partial_{y}( \psi^\dagger \partial_{x}\psi )]
\end{eqnarray}
The last integral involves full derivatives and can be readily done using the boundary conditions (\ref{twistp}). The result depends only on the
"Abelian vector potentials" $\alpha(x,y), \beta(x,y)$ and coincides with the flux of the corresponding auxiliary  magnetic field, so
\beq
N_{CS} \ =\ 1/N
\eeq
in this case. 
It is clear that $U^{p}$ gives rise to a configuration with $N_{CS} = p/N$.

We have considered the case when $P$ and $Q$ commute to the primitive root 
of unity $\epsilon$. The construction generalizes straightforwardly to the 
case of
$QP = e^{2\pi i k / N}PQ$ when $k$ and $N$ are coprime. Along the same 
lines as above,  one easily obtains $U(\vec{x})$ 
satisfying  Eq. (\ref{twistu}).
The corresponding CS number is equal to $k / N$. Powers of U complemented by 
instanton
 shifts will generate
all values of $N_{CS}$ that are multiple integers of   $1/N$ .

Now suppose that $N$ is not prime and $k$ is its divisor: $N=kM$.
Take $\omega_1 =\epsilon^k=e^{2\pi i k / N}=e^{2\pi i/ M}$. 
This  element  generates  a subgroup 
$Z_M$ of the center group $Z_N$. In this case, there are moduli of 
the Heisenberg pairs \cite{schw,ks}.
The centralizer of a pair is now a continuous subgroup of $SU(N)$. 
There are some special  pairs with centralizer $SU(k)$,
\footnote{Their explicit form is
$$
P  = \ e^{2\pi i l/k} {\rm diag} (p, \ldots, p) , \ \ \ \ \ \ \ \ 
Q = \ e^{2\pi i s/k} {\rm diag} (q, \ldots, q) \ ,
$$
$p, q \in SU(M),\ qp = e^{2\pi i /M}pq,\ \ l,s = 1,\ldots, k$. Note that 
the pairs characterized by  different  $l,s$ are inequivalent to each other by 
conjugation. } while the centralizer 
of a generic pair
is $[U(1)]^{k-1}$ , the maximal torus $T$ in   $SU(k)$ .
Let us consider the 
Heisenberg  pair  $P$ and $Q^{k}$,  with
$P$ and $Q$ defined in Eq. (\ref{PQ}). This pair is of generic type,
so the moduli space of the corresponding flat connections is  $[U(1)]^{k-1}$.
In terms of the matrix $U$ in Eq. (\ref{flat}) these moduli show up
in the boundary conditions which now read as follows:
\begin{eqnarray}
\label{twistum}
U(x+1)\ =\ e^{iT_{1}}PU(x)P^{-1}\nonumber\\
U(y+1)\ =\ e^{iT_{2}}Q^k U(y)Q^{-k}\\
U(z+1)\ =\ \epsilon e^{iT_{3}}U(z),\nonumber
\end{eqnarray}
where $\exp\{iT_{j}\}$ lie on the  torus $T$.
The moduli $T_{j}$ are easily taken into account by substituting
\beq
\label{moduli}
U(x) \ =\  e^{i  \vec{x} \vec{T}} \tilde U(\vec{x})\ , 
\eeq 
so that  $\tilde U(\vec{x})$ satisfy the boundary conditions (\ref{twistum})  without the
factors   $\exp\{iT_{j}\}$.

To find $\tilde U(\vec{x})$ ,  we can use the same ansatz as before [see Eqs. 
 (\ref{ansatz}), (\ref{fund}), (\ref{norm})] 
and everything goes through in a parallel way with the the only change
that, to compensate for the commutant of $P$ and $Q^k$, we should take  $\alpha(x,y) = 2\pi i ky/N$ so that  
 the flux of the auxiliary
magnetic field is now $ k / N$.
Eq. (\ref{pj}) now reads
\beq
\label{pjnew}
\psi_{l+kj}(x,y)\ =\ \psi_{l}(x,y+j)\ , \ \  \\
l \ = \ 1, \ldots, k \ ; \ \ \ \ \ j = 1, \ldots , M-1 \ .
\eeq
It expresses 
all components of $\psi$ in terms of the first $k$ components. We can safely assume
that only one of $ \psi_{l}$, \ $l = 1, \ldots, k$ , say, the one with $l=1$, 
is different from zero. The functions
$\tilde U$ obtained with  more general assumptions  obey the same boundary conditions
and thus  are all gauge equivalent. The conditions (\ref{pjnew}) imply that
$\psi_1(x,y)$ and all its ``descendants'' are   
 periodic in $y$ with period $M$. On the other hand,
\beq
\label{p1xnew}
\psi_{1}(x+1,y)\ =\ e^{2\pi iy/M} \psi_{1}(x,y) \ .
\eeq
The conditions (\ref{pjnew}), (\ref{p1xnew}) have exactly the same form as Eqs.
(\ref{pj}), (\ref{p1x}), only $N$ is substituted by $M$. A 
set of functions $\psi$ satisfying these conditions  can be 
chosen as
\beq
\label{psinew}
\psi_{1+kj}(x,y)\ =\ {\cal N}(x,y)
e^{-\pi (\frac{y}{M})^{2}+2\pi i x\frac{y}{M}}
\Theta_{j/M, 0}  \left(x+ i \frac{y}{M} , i \right)\ 
\eeq
with $j = 0,\ldots, M-1$. The
CS number is computed as easy as before and is equal to 1/M. The connections with
$U^{(p)}(x,y) =  [\tilde U (x,y)]^p$ (they satisfy the boundary conditions
(\ref{twistum}) with $T_i = 0$ and $\epsilon \to \epsilon^p$)  have the 
Chern--Simons number $p/M$.  
  
\section{Flat connections for exceptional triples.}
As was explained in the Introduction, exceptional  triples in higher
orthogonal and exceptional groups are intimately
connected with Heisenberg pairs for unitary groups. We will explain now how
the corresponding periodic connections are build up in terms of the twisted   
connections for $SU(N)$. 

Let $G$ be a simple connected simply connected Lie group. 
Pick up a generic exceptional $\Omega_1$ whose centralizer is a product
$H \ = SU(N_{1}) \times SU(N_{2}) \times \ldots $
 factorized over a subgroup of its center. The liftings
of $\Omega_{2,3}$ in $H $ 
are Heisenberg pairs in each component $SU(N_i)$. Let us find out how
$\Omega_1$ is embedded into $H$. 
A priori, $\Omega_{1}$ could be an element of the center
$Z_{N_{1}}  \times Z_{N_{2}} \times  \ldots$ of $H$. In fact, the known
explicit form
of  $\Omega_{1}$  allows one to conclude that
it is unity (the trivial element of the center) in all
$SU(N_{j})$ components but the component $SU_{\theta}(N_{\theta})$  
which contains  the  coroot $\theta^{\lor}$ corresponding to the highest
root $\theta$  as a generator in its maximal torus. 

Indeed, as follows from Theorem 1 of 
Ref.\cite{ks}, any exceptional element can be conjugated to
 \be
\label{except}
\Omega_1 \ =\ \exp\left\{ 2\pi i \sum_{j=1}^r s_j \omega_j \right\} \ ,
 \ee
where the sum runs over all nodes of the Dynkin diagram of the corresponding group, $r$ is the rank of the group, $\omega_j$ are the fundamental coweights, i.e. the elements of the
Cartan subalgebra commuting with all simple root vectors but one,
$[\omega_j, T_{\alpha_k}] = 
\delta_{jk} T_{\alpha_k} $, and the real numbers
$s_i$ (so called {\it Kac coordinates}) have the following properties:
 \begin{itemize}
\item $s_j \geq 0$.
\item $\sum_{j=1}^r a_js_j \ = 1$, where $a_j$ are {\it Dynkin labels}, or the integer coefficients of expansion of the highest root
$\theta$  over simple roots, $\theta \ =  \ \sum_j a_j \alpha_j $,
of the corresponding node. 
\item The greatest common divisor $m$ of all {\it dual} Dynkin labels
$a_j^\lor = a_j \langle \alpha_j, \alpha_j \rangle /2$ 
dwelling on the nodes with  $s_j \neq 0$ is nontrivial $m > 1$. The dual Dynkin labels are the coefficients of expansion of the coroot $\theta^\lor$ corresponding to the highest root $\theta$ over the simple coroots $\alpha^\lor$. For simply laced groups $a^\lor_j = a_j$.  The integer $m$ is an important characteristic  of the corresponding exceptional triple and can be called its {\it order}.
 \end{itemize}

To find the (semi--simple part of) the centralizer, one has to
{\it (i)} Consider the extended Dynkin diagram of the group including the
simple roots $\alpha_j$ and the root $\alpha_0 = -\theta $ and to cross out all the nodes for which $s_j \neq 0$.
Generically, we are left with a product $H$ of unitary groups. 
\ {\it (ii)} Factorize it over $Z_m$ embedded in a certain way in the center of $H$. 

Now, $\omega_j$ entering the sum in Eq. (\ref{except}) 
commute with all root vectors 
corresponding to the nodes not entering the sum. The relation
$[\omega_j, T_{\alpha_0}]    = -a_j T_{\alpha_0}$ holds.
Taking into account this and the condition  $\sum_{j} a_js_j \ = 1$, we see that $g = \sum_{j} s_j \omega_j $ commutes with all components $SU(N_i)$ not involving the root $\alpha_0$ and  that
$[g, T_{\alpha_0} ] = - T_{\alpha_0} $. Therefore, $g$ represents
(up to irrelevant sign) the fundamental coweight $\omega_0$ associated with the root 
$\alpha_0$ in  $SU_{\theta}(N_{\theta})$. This coweight is conjugate to the matrix $T_0$ in Eq. (\ref{t8})   and one easily sees that  the element (\ref{except}) is trivial from the viewpoint of all 
subgroups $SU(N_j) \subset H$ but the group $SU_{\theta}(N_{\theta})$, where it represents
a generating  element of the center. 
\footnote{To be quite precise, one should not say `` $\Omega_1$ {\it is} trivial, etc ", but rather that it {\it can be chosen} as such.
Indeed, the true centralizer is not $H$, but $H$ factorized over a
nontrivial subgroup of its center, and different liftings of $\Omega_i$ in $H$ are possible. The final result does not depend, of course, on the choice of the lifting.  See the discussion of the $Spin(7)$ example below.}
By inspection of different cases \cite{ks}, one can observe that $N_\theta$ always 
coincides with the order $m$ of the triple.

Let now ${\tilde \Omega_{j}}$  stand for  $\Omega_{j}$
projected onto $SU_{\theta}(m)$.
Take   the matrix $U(x,y,z)$ obeying the 
boundary conditions 
\begin{eqnarray}
\label{twistu1}
U(x+1)={\tilde \Omega_{2}}U(x){\tilde \Omega_{2}}^{-1}\nonumber\\
U(y+1)={\tilde \Omega_{3}}U(y){\tilde \Omega_{3}}^{-1}\\
U(z+1)={\tilde \Omega_{1}} U(z).\nonumber
\end{eqnarray}
Take the explicit expression for this matrix from the previous section
and lift it up to the group $G$. 
The last subtlety is that the gauge field corresponding to 
such lifting $U_G$  is not periodic.  Instead,
\begin{eqnarray}
\label{twist1}
A_i(x+1)=\Omega_{2}A_i(x)\Omega_{2}^{-1}\ ,\nonumber\\
A_i(y+1)=\Omega_{3}A_i(y)\Omega_{3}^{-1} \ ,\\
A_i(z+1)=A_i(z)\ .\nonumber
\end{eqnarray}
However, this is curable, since $\Omega_{2}$ and $\Omega_{3}$
can be simultaneously conjugated to the maximal torus in $G$,
\beq
\label{tor}
\Omega_{2}=e^{ia}, \; \Omega_{3}=e^{ib}, \; [a,b]=0\ .
\eeq

Thus, taking instead of $U_G$ the element $\tilde U_G$
\beq
\label{raskrut}
{\tilde U}_G(x,y,z)=U_G(x,y,z)e^{i(ax+by)},
\eeq
one finally obtains the periodical zero curvature 
gauge field corresponding to the exceptional triple of holonomies
$\Omega_{1},\, \Omega_{2}, \, \Omega_{3}$.

Let us now comment on the CS number of the flat connection obtained in this way. 
First of all, one can verify by direct computation that CS does not change
under the transformation  (\ref{raskrut}), so that one can work with 
$U_G$, not with  $\tilde U_G$.
The total CS is equal to the sum of the CS numbers in each component  
$SU(N_i) \subset H$ weighted with certain integer factors $n_i$ reflecting a particular way of how $SU(N_j)$ is embedded in the large group $G$. For simply laced groups $n_j = 1$ in all cases. For the groups $Spin(2r + 1)$  we have generically 
$H = SU_{\alpha_1}(2) \times SU_{\alpha_2}(2) \times SU_{\theta}(2)$, where one of the roots $\alpha_{1,2}$ is short. If $\alpha_1$ is short, $n_1 = 2$ 
(the corresponding coroot is long and the contribution to the
integral  in Eq. (\ref{cs}) is 
 proportional to $\langle \alpha_1^\lor, \alpha_1^\lor \rangle 
= 2 \langle \alpha_2^\lor, \alpha_2^\lor \rangle $)
and $n_2 = n_\theta = 1$ . For $G_2$, $H = SU_{\alpha}(2) \times
SU_{\theta}(2)$, where $\alpha$ is the short root with 
$n_\alpha = 3$, while $n_\theta = 1$. Thus, $n_\theta = 1$ in all cases (actually, it is
a theorem that the highest root $\theta$ is always long and the corresponding coroot 
$\theta^\lor$ is always short). 

Now, for the components not involving $\theta$, $\Omega_1$ is represented by the unity and it is known
\cite{thooft,witten} that the gauge field configurations constructed via matrices $U$ satisfying the boundary conditions (\ref{twistu1}) with $\tilde\Omega_1 = 1$ have instanton nature and an integer Chern--Simons number, which is as good as zero for our purposes. In fact, one can just take $U =1$ in all these subgroups and forget about them. 

 Speaking of 
the component $SU_{\theta}(m)$,  we have seen that $\Omega_{1}$ is represented there by a generating element of the center $Z_m$. The calculation of the CS number boils down to the
one from the previous section. 
$N_{CS}$ depends on the commutant of $\tilde \Omega_2$ and $\tilde \Omega_3$ and is an integer multiple of $1/m$.

To make things absolutely clear, let us  illustrate this general 
construction in the case of $Spin(7)$ group, 
the smallest orthogonal group where one meets
an exceptional triple. 

\begin{figure}
\grpicture{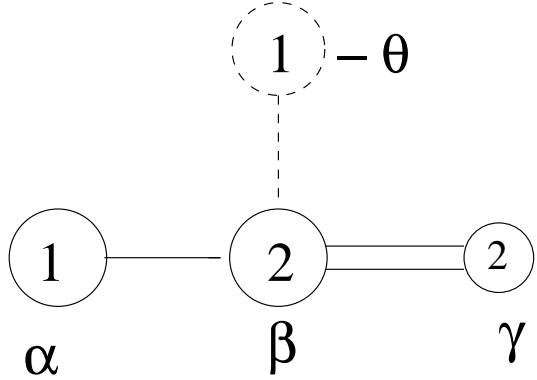}
\caption{Dynkin diagram for $Spin(7)$ with its Dynkin labels.}
\label{Dyn7}
\end{figure}

Consider the extended Dynkin diagram of $Spin(7)$ (Fig. \ref{Dyn7}). The simple coroots are $\alpha^\lor = e_1 - e_2,\ \beta^\lor = e_2 - e_3,\ 
\gamma^\lor = 2e_3$, where $e_{1,2,3}  $ are the generators of rotation in 3 independent planes. The highest coroot is $\theta^\lor = \alpha^\lor + 2\beta^\lor + \gamma^\lor = e_1 + e_2$ (remember that $a_\gamma^\lor = a_\gamma/2 = 1$) and it coincides in this case with the fundamental coweight $\omega_\beta$. 
The first element of the  triple is  
\beq
\label{o1}
\Omega_{1}=e^{i \pi \omega_{\beta}} = \ e^{i\pi\theta^\lor}\ .
\eeq
The centralizer of such $\Omega_{1}$ in $Spin(7)$
is $[SU_\alpha(2) \times SU_\gamma(2) \times SU_\theta(2)]/Z_2$,
where $Z_{2}$ is the diagonal subgroup in the center group
$Z_{2}^\alpha \times Z_{2}^\gamma \times Z_{2}^\theta$. Then the Heisenberg
pair in  $SU_\alpha(2) \times SU_\gamma(2) \times SU_\theta(2)$ can be taken as
follows
\begin{eqnarray}
\label{o23}
\Omega_{2}=(i\sigma_{3}, i \sigma_{3}, i\sigma_{3}) \ ,\nonumber\\
\Omega_{3}=(i\sigma_{1}, i\sigma_{1}, i\sigma_{1})\ ,
\end{eqnarray}
where $\sigma_i$ are the standard Pauli matrices. The expression (\ref{o1}) is rewritten as
$\Omega_1 = (1,1,-1) \equiv (-1,-1,1)$ in the direct product notations. Choosing the first possibility,
lifting the twisted connection in $SU_\theta(2)$ up to $Spin(7)$, and performing the gauge transformation (\ref{raskrut}), we obtain an interesting periodic $Spin(7)$ connection with $N_{CS} = 1/2$.  (
If  choosing the second possibility, we obtain the same result up to an irrelevant integer. Indeed, $n_\gamma = 2$ and the twisted connection in the group 
$SU_\gamma(2)$ provides an integer contribution to the integral (\ref{cs}). The twisted connection of the subgroup  $SU_\alpha(2)$ provides the contribution $1/2$ to $N_{CS}$. )

Consider now the connections whose holonomies form {\it Heisenberg triples} (``almost commuting" triples by the
terminology of Ref. \cite{bfm}), i.e. triples $\Omega_i$ of group elements such that 
$\Omega_i \Omega_j = c_{ij} \Omega_j\Omega_i $, with $c_{ij}  $ being nontrivial elements of 
the center of the group , for nonunitary groups $G$. They  also can be expressed in terms of twisted unitary 
group connections constructed in the previous section. We will describe the simplest such 
example with the group $Sp(4)$ and perform the calculation of Witten index for the 
${\cal N} = 1$ supersymmetric gauge theory based on this group.

$Sp(4)$ is a subgroup of $SU(4)$ containing all unitary $4 \times 4$ matrices $U$ satisfying the condition 
$U^T I U = I$, where $I$ is an antisymmetric symplectic matrix, which can be chosen in the form 
$I = {\rm diag} (i\sigma_2, i\sigma_2) \equiv \ \vec{1} \otimes i\sigma_2$. The group $Sp(4)$ has center $Z_2$
 with nontrivial element diag$(-{\bf 1},-{\bf 1}) \equiv - \vec{1} \otimes \vec{1}$. 

 Heisenberg triples are constructed by studying first the Heisenberg pairs in $Sp(4)$  \cite{bfm}.  
Any pair $P,Q$ with $QP = -PQ$  can be conjugated to
\be
\label{parSp4}
P \ =\ e^{i\alpha \sigma_2} \otimes i\sigma_3,\ \ \ \ \ 
 Q \ =\ e^{i\beta \sigma_2} \otimes i\sigma_{\bf 1} \ ,
  \ee
where the angular variables $\alpha, \beta$ are moduli. 
The centralizer of a generic pair $P,Q$ in Eq. (\ref{parSp4} ) is $U(1)$, the corresponding matrices being given by
  \be
  \label{cenSp4}
 S\ =\  e^{i\gamma\sigma_2}  \otimes {\bf 1} \ .
  \ee
 The Heisenberg triples $P,Q,S$ give rise to the ``principal'' component ( i.e. the component
involving the perturbative vacuum $A_i = 0$) of the moduli space of classical
 vacua. They all have zero (or integer) Chern--Simons number. As the rank of the centralizer  of the pair $P,Q$ is
 1, we obtain $r+1 = 2$ quantum vacuum states.   
 
   There are, however, some special points in the moduli space of the pairs where the centralizer is larger.  
There are three points:  $\alpha = 0, \beta = \pi/2$,  $\alpha = \pi/2, \beta = 0$, and  $\alpha = \pi/2, \beta = \pi/2$ where the centralizer is $SU(2)$ \cite{ks,keur3}, but, as $SU(2)$ is connected and contains 
unity, we are still in the principal vacuum sector.  
If $\alpha=\beta=0$ , the  centralizer is $O(2)$ and involves besides (\ref{cenSp4}) the disconnected component 
$ e^{i\gamma\sigma_2}\sigma_3  \otimes {\bf 1} $, which is equivalent by conjugation to the element
$\sigma_3 \otimes {\bf 1} \ =$ diag$({\bf 1}, -{\bf 1})$
($P,Q$ are invariant under such conjugation) \cite{bfm,witten3}. Now, the unique up to conjugation exceptional Heisenberg triple of holonomies 
 \be
 \label{tripSp4}
P \ =\ {\bf 1} \otimes i\sigma_3,\ \ \ \ \ 
 Q \ =\ {\bf 1} \otimes i\sigma_1 \ , \ \ \ \ S \ =\ \sigma_3 \otimes {\bf 1} 
 \ee
 defines the unique up to conjugation  interesting twisted connection in $Sp(4)$ , which is further promoted to the supersymmetric quantum vacuum state.
All together, we have $2+1 = 3$ vacuum states which coincides with
the counting $r_{Sp(4)} + 1$ based on the analysis of periodic connections.  The interesting
 connection based on the triple (\ref{tripSp4}) is obtained from a matrix $U$ satisfying 
  \begin{eqnarray}
   \label{twistSp4}
 U(x+1)=PU(x)P^{-1}\nonumber\\
 U(y+1)=QU(y)Q^{-1}\\
 U(z+1)= S \cdot U(z)\ ,\nonumber
  \end{eqnarray}
These boundary conditions can be easily satisfied with the ansatz $U(\vec{x}) = {\rm diag} ({\bf 1}, u(\vec{x}))$, where $u$ is the $SU(2)$ matrix satisfying the twisted boundary conditions (\ref{twistu}) and which was found before. The corresponding gauge field configuration has Chern--Simons number $1/2$.

\section{Discussion}

We have described explicitly classical vacua in Yang-Mills on $T^{3}$, that is
we have
constructed analytic expressions for twisted flat connections in unitary groups 
and have shown that, by a proper embedding, they define also interesting periodic 
and also twisted flat connections in more complicated groups. 
The explicit form of the gauge fields allowed us to verify the corresponding
fractional CS charges \cite{bfm} by direct computation.

With our explicit construction in hand, the next challenge is to find 
 Euclidean 4--dimensional gauge field solutions to the classical equations 
of motion which interpolate between different vacua in complicated groups. 

It is not difficult to present a heuristic reasoning in favor of  existence of such solutions on 
$T^3\times R$   \cite{krs}. Consider a
flat connection $A_i^{\rm flat}(\vec{x})$ belonging to a topologically nontrivial  component 
not involving the configuration $A_i = 0$. Multiply it by a function $f(\tau)$ of Euclidean time 
such that $f(\infty) = 1$ and $f(-\infty) =0$. The interpolating configuration 
$f(\tau)A_i(\vec{x})$ has a nonzero field strength and a nonzero Euclidean action. 
The latter is not necessarily minimal, but it is clear that exploring the directions in 
the functional space which make the action lower, we will finally find a configuration 
with minimal action, i.e. the classical solution.

Of course, this  does not tell us  what {\it is} this solution and, 
in particular, what is its action. Also, this reasoning does not exclude 
that the configuration minimizing the action  
becomes singular (cf. the problem of finding instanton solutions of unit charge on $T^4$, 
where such a "singularization" occurs, indeed \cite{Baal}).

These issues can be  clarified, bearing in mind  the fact that  flat connections  for complicated 
groups are constructed out of twisted flat connections for $SU(N)$. 
It was proven earlier \cite{Todorov} that   Euclidean solutions 
with fractional instanton number $\nu = 1/N, \ldots$ which interpolate between the perturbative vacuum
$A_i = 0$ and a nontrivial twisted vacuum exist. They are not singular and satisfy also self--duality
equations $F_{\mu\nu} = \pm \tilde  F_{\mu\nu}$, which means that their action is equal to 
$8\pi^2/N$. Such solutions were studied numerically in Refs.  \cite{simulation}.

Performing a proper embedding    and perhaps a gauge transformation like in 
Eq. (\ref{raskrut}), we are
led to the existence of  self--dual solutions 
on $T^3 \times R$ that interpolate between  flat 
connections belonging to different topological classes as discussed above
 for an arbitrary group.
The instanton numbers  of these solutions are multiple integers of $1/m$,
 or rather   of the differences $1/m - 1/m'$, where $m$ and $m'$ are different 
admissible 
orders of exceptional triples. For example, for $E_8$ the instanton number $\nu$ 
can be as 
small as $\nu = 1/5 - 1/6 = 1/30$. The action of these solutions is $8\pi^2 \nu$ in
all cases.

Hopefully, our explicit constructions will help in finding an educated
guess about analytic form of those solutions.
Another interesting problem is to study possible dualities
relating the classical vacua with different gauge groups
and different boundary conditions. Given the explicit
formulae containing $\Theta$ functions with rational characteristics, 
one could expect to see many such dualities associated with the duality
relations between $\Theta$ functions.

We are grateful to P. van Baal, A. Keurentjes, and A.Rosly for useful discussions.  
K.S. thanks SUBATECH, Nantes and Math. Department of Nantes University,
where this work was done, for  support and hospitality. The work of K.S. was 
also supported by CRDF RP1-2108, INTAS 97-0103, and RFBR grant for 
Scientific Schools 00-15-96562.

\end{document}